\documentstyle[12pt]{article}
\def\fun#1#2{\lower3.6pt\vbox{\baselineskip0pt\lineskip.9pt
\ialign{$\mathsurround=0pt#1\hfil##\hfil$\crcr#2\crcr\sim\crcr}}}
\newcommand{\be}{\begin{eqnarray}}
\newcommand{\ee}{\end{eqnarray}}
\newcommand{\bd}{\begin{displaymath}}
\newcommand{\ed}{\end{displaymath}}
\newcommand{\ba}{\begin{array}}
\newcommand{\ea}{\end{array}}
\newcommand{\bc}{\begin{center}}
\newcommand{\ec}{\end{center}}

\input{epsfig.sty}

\begin{document}

\Huge{\noindent{Istituto\\Nazionale\\Fisica\\Nucleare}}

\vspace{-3.9cm}

\Large{\rightline{Sezione SANIT\`{A}}}
\normalsize{}
\rightline{Istituto Superiore di Sanit\`{a}}
\rightline{Viale Regina Elena 299}
\rightline{I-00161 Roma, Italy}

\vspace{0.65cm}

\rightline{INFN-ISS 95/18}
\rightline{December 1995}

\vspace{1cm}

\bc
\large{\bf HEAVY-TO-HEAVY AND HEAVY-TO-LIGHT \\ WEAK DECAY FORM FACTORS
IN THE LIGHT-FRONT \\ APPROACH: THE EXCLUSIVE $ 0^- $ TO $ 0^- $ CASE
\footnote{{\normalsize \bf ~ To appear in Soviet Journal of Nuclear
Physics (1996).}}}

\vspace{1cm}

{\bf N.B. Demchuk$^{(a)}$, I.L. Grach$^{(a)}$, I.M. Narodetskii$^{(a)}$,
S. Simula$^{(b)}$}

\vspace{1cm}

\normalsize{$^{(a)}$Institute for Theoretical and Experimental Physics,
\\ 117259 Moscow, Russia \\ $^{(b)}$Istituto Nazionale di Fisica
Nucleare, Sezione Sanit\`{a}, \\ Viale Regina Elena 299, I-00161 Roma,
Italy}

\ec

\vspace{1cm}

\begin{abstract}

Weak transition form factors among heavy pseudoscalar mesons are
investigated within a relativistic quark model formulated on the
light-front. It is shown that the light-front result derived in the
time-like region for the matrix elements of the plus component of the
weak vector current coincides with the spectator pole term of the quark
triangle diagram. For the first time, the dependence of the form factors
on the squared four-momentum transfer $q^2$ is calculated in the whole
accessible kinematical region $0 \leq q^2 \leq q^2_{max}$. For the
numerical investigations of the semileptonic $B \to D \ell \nu_{\ell}$,
$B \to \pi \ell \nu_{\ell}$, $D \to K \ell \nu_{\ell}$ and $D \to \pi
\ell \nu_{\ell}$, the equal-time wave functions corresponding to the
updated version of the ISGW model are adopted. Our results for the form
factors and the decay rates are presented and compared with available
experimental data and predictions of different approaches. Moreover,
the $K^0_L \to \pi^{\pm} ~ e^{\mp} ~ \nu_e$ decay is briefly discussed.
Our approach is consistent with experimental data.

\end{abstract}

\newpage

\section{Introduction}

Semileptonic heavy-quark decays play an important role for the
understanding of weak and strong interactions. These decays
proceed via the spectator mechanism, in which the heavy-quark decays
into another heavy or a light quark by emitting a W-boson, which
materializes into a lepton pair. The decay amplitudes are given by the
product of the leptonic and hadronic $V - A$ currents. Since only two
quarks are in the final state, no interfering diagrams and no effects
from the final state interactions should be taken into account, at
variance with the case of non-leptonic decays. The matrix elements of
the hadronic current are determined by the bound state properties of the
initial and final hadrons and, therefore, they can provide relevant
information on the internal structure of heavy hadrons. The knowledge
of the weak hadron form factors, combined with measurements, allows
to determine some fundamental quantities of the Standard Model, the
Cabibbo-Kobayashi-Maskawa (CKM) parameters. The knowledge of the hadron
matrix elements is also of interest for estimating the non-leptonic
decays of heavy hadrons. A theoretical challenge consists in the
appropriate description of the non-perturbative aspects of the strong
interaction. The non-perturbative approaches, commonly applied in this
field, include the Heavy Quark Effective Theory (HQET), lattice QCD
calculations, the QCD sum rules (SR) and constituent quark models (CQM).

In order to calculate the decay rates and to determine the CKM
parameters from experiment, it is necessary to know both the
normalization and the dependence of the hadron form factors upon the
squared four-momentum transfer $q^2$. There are few lattice QCD
calculations of the $B \to D$ transitions in the limit of infinite $b$
and $c$ quark masses [1], while there are various QCD SR predictions
[2-5], which, however, do not cover the full range of $q^2$. It should
be stressed that even within the CQM the evaluation of the
$q^2$-dependence of the form factors in the whole kinematical region
accessible in semileptonic decays have not yet been performed. Usually,
the form factors are calculated at a fixed value of $q^2$ appropriate
for the specific model and, then, extrapolated to the full range of
$q^2$. In the original BSW approach [6] the form factors are calculated
at $q^2 = 0$ and extrapolated to $q^2 > 0$ using a monopole form factor
$(1 - q^2 / M_{pole}^2)^{-1}$, where $M_{pole}$ is the mass of the
lowest-lying meson resonance relevant in the given decay channel.
Within the ISGW model [7,8] the form factors are calculated at the
point of zero recoil (i.e., $q^2 = q_{max}^2$) and, then, extrapolated
to $q^2 = 0$ assuming an exponential-like behaviour of the form factor.
A relativistic CQM model due to Jaus [9] uses the light-front (LF)
formalism to compute the form factors for space-like values of the
momentum transfer (i.e., $q^2 \leq 0$). Then, the form factors are
extrapolated to the time-like region $q^2 > 0$ assuming a particular
two-parameter formula, which reproduces the values of the form factors
and their first two derivatives at $q^2 = 0$.

In this paper, we adopt a relativistic CQM and, for the first time, we
present the results of the calculation of the $q^2$ dependence of weak
transition form factors among pseudoscalar mesons in the whole
accessible kinematical region. As in ref. [9], our approach is based on
the LF formalism, which allows a proper treatment of the effects of
both relativistic composition of quark spins and center-of-mass recoil.
However, instead of the Breit frame, which is appropriate only for
space-like values of $q^2$, our calculations are performed in a
reference frame where the momentum transfer is purely longitudinal,
which is appropriate for time-like values of $q^2$. The relevance of
such a frame for the calculation of the Isgur-Wise function has been
stressed in ref. [10].

The plan of the paper is as follows. In Section 2 we describe our
approach for the calculation of the weak pseudoscalar meson form
factors. We evaluate the triangle diagram for the weak vector current,
where the integration over the light-front energy for the plus
component of the current is the essential step. Then, we compare this
result with the one obtained within the Hamiltonian light-front
dynamics. We show that the latter coincides with the contribution of
the spectator pole in the quark triangle diagram. In Section 3 our
results for the form factors and the exclusive decay rates of the $B \to
D \ell \nu_{\ell}$, $B \to \pi \ell \nu_{\ell}$, $D \to K \ell
\nu_{\ell}$ and $D \to \pi \ell \nu_{\ell}$ weak decays, are presented
and compared with the experimental data as well as with results of
different approaches. Moreover, the $K^0_L \to \pi^{\pm} ~ e^{\mp} ~
\nu_e$ decay is briefly discussed. Finally, our conclusions are
summarized in Sect. 4.

\section{Weak decay vector form factor}

\subsection{Kinematics}

Here below, we denote by $P_1, P_2$ and $M_1, M_2$ the 4-momenta and
masses of the parent and daughter mesons, respectively. The
four-momentum transfer $q$ is given by $q = P_1 - P_2$ and the
invariant $y$ is defined as
 \be
    \label{1}
     y = {P_2^+ \over P_1^+} = 1 - {q^+ \over P_1^+}
 \ee
We work in a frame where the momentum transfer is such that ${\bf
q_{\bot}}=0$. Thus, it can be easily checked that
 \be
    \label{2}
    q^2 = (1 - y) \left( M_1^2 - \frac{M_2^2}{y} \right)
 \ee
yielding two solutions for the variable $y$, viz.
\be
   \label{3}
   y_{1, 2} = \zeta ~ (\eta \pm \sqrt{\eta^2 - 1})
 \ee
where $\zeta \equiv M_2/M_1$ and the "velocity transfer" $\eta$ is
defined as $\eta = U_1 \cdot U_2 = P_1 \cdot P_2 / (M_1 M_2)$, with
$U_1$ and $U_2$ being the four-velocities of the parent and daughter
mesons. The two signs in eq. (3) correspond to whether one chooses the
3-momentum ${\bf p_2}$ of the final meson to be in the positive or
negative direction of the 3-axis. As is well known, the relation
between the kinematical variables $\eta$ and $q^2$ is given by
 \be
    \label{4}
    q^2 = M_1^2 (1 + \zeta^2 - 2\zeta \eta)
 \ee
At the point of zero recoil, i.e. $q^2 = q_{max}^2 = (M_1 - M_2)^2$
corresponding to $\eta = 1$, one has $y_{1, 2}(q_{max}^2) = \zeta$,
while at $q^2 = 0$, corresponding to $\eta = \eta_{max} \equiv {1 \over
2} (\zeta + {1 \over \zeta})$, one gets $y_1(0) = 1$ and $y_2(0) =
\zeta^2$.

\indent In this paper we are interested in the weak decay of a
pseudoscalar meson into another pseudoscalar meson. We define the form
factors of the $P_1(Q_1 \bar{q}) \to P_2(Q_2 \bar{q})$ transition
between two pseudoscalar mesons in the usual way, viz.
 \be
    \label{6}
    <P_2| \bar{Q}_2 \gamma_{\mu} Q_1 |P_1> & = & \sqrt{M_1 M_2} ~
    \left[ h_+(q^2) ~ (U_1 + U_2)_{\mu} + h_-(q^2) ~ (U_1 - U_2)_{\mu}
    \right] \\ \label{7}
    & = & f_+(q^2) ~ (P_1 + P_2)_{\mu} + f_-(q^2) ~ (P_1 - P_2)_{\mu}
 \ee
where $J_{\mu} \equiv \bar{Q}_2 \gamma_{\mu} Q_1$ is the weak vector
current. Let us remind that the term containing the form factor
$f_-(q^2)$ in eq. (\ref{7}) yields a contribution to the semileptonic
decay rate proportional to the lepton mass and, therefore, it does not
contribute to the transition amplitude, except in case of the heavy
$\tau$ lepton and $K_{\mu 3}$ decay. The relationship between the two
sets of form factors is given by
 \be
    \label{8}
    f_{\pm}(q^2) = \frac{1}{2 \sqrt{M_1 M_2}} \left( (M_1 + M_2) ~
    h_{\pm}(q^2) - (M_1 - M_2) ~ h_{\mp}(q^2) \right)
 \ee
Another set of form factors can be introduced, viz.
 \be
    \label{9}
    <P_2| \bar{Q}_2 \gamma_{\mu} Q_1 |P_1> = F_1(q^2) \left( P_1 + P_2 -
    \frac{M^2_1 - M^2_2}{q^2} q \right)_{\mu} + F_0(q^2) \frac{M_1^2 -
    M_2^2}{q^2} q_{\mu}
 \ee
with
 \be
    \label{10}
    F_1(q^2) & = & f_+(q^2) \\ \nonumber
    F_0(q^2) & = & f_+(q^2) + \frac{q^2}{M_1^2 - M_2^2} ~ f_-(q^2)
 \ee
The physical meaning of these new form factors is clear in the helicity
basis, where the form factors $F_1$ and $F_0$ are associated with the
transition amplitudes corresponding to the exchange of a vector $(1^-)$
and a scalar $(0^+)$ boson in the t-channel. However, the masses of
heavy scalar mesons are poorly known. Therefore, in what follows, the
form factors $f_{\pm}(q^2)$ within the assumption of pole dominance are
constructed in a much simpler way as follows
 \be
    \label{11}
    f_{\pm}(q^2) = {f_{\pm}(0) \over 1 - q^2 / M_{1^-}^2}
 \ee
where $M_{1^-}$ denotes the mass of the lowest-lying vector $\bar{Q}_2
Q_1$ meson. This simple picture has clear limitations that will be
addressed in Section 3.

Neglecting the lepton masses, the rate $\Gamma$ for semileptonic decays
reads as
 \be
    \label{12}
    \Gamma = \frac{16}{3} ~ \Gamma_0 ~ |V_{Q_1 Q_2}|^2 ~ \zeta^4 ~
    \int_1^{{1 \over 2} (\zeta + {1 \over \zeta})} d\eta ~ (\eta^2 -
    1)^{3/2} ~ f_+^2(\eta)
 \ee
where $\Gamma_0 = G_F^2 M_1^5 / (4\pi)^3$ and $V_{Q_1 Q_2}$ is the
relevant CKM matrix element.

\subsection{The Feynman triangle diagram}

The LF wave function is defined on the plane $x^0 + x^3 = 0$ and can be
obtained from the Bethe-Salpeter amplitude by performing the integration
over the LF "energy" $k^- = k_0 - k_3$ (see, e.g., ref. [9]). As a
matter of fact, in ref. [11] the pion charge form factor has been
evaluated starting from the quark triangle diagram in a Breit frame
where $q+ \equiv q_0 + q_3 = 0$, and the integration over $k^-$ in the
loop integral allowed the identification of the LF wave function.
Moreover, it is known [12,13] that, when $q^+ = 0$, the use of the
so-called "good" component of the current $J^+ \equiv J^0 + J^3 $
allows to suppress the effects of quark pair creation from the vacuum.
In this way the quark triangle diagram is equivalent to the result
which can be obtained within the Hamiltonian LF dynamics [14]. In this
subsection we evaluate the quark-triangle diagram for the weak
vector current, using a frame where the momentum transfer is purely
longitudinal, i.e. ${\bf q}_{\perp} = 0$ and $q^+ > 0$, as it is
appropriate for time-like values of $q^2$. We find two contributions,
which will be referred to as the partonic and non-partonic ones. In the
next subsection we will show that the partonic term coincides with the
result obtained using the Hamiltonian LF approach. The non-partonic
contribution corresponds to the quark pair creation diagram and its
calculation will not be addressed in this paper. An estimate made in
ref. [10] shows that the non-partonic contribution may be expected to
be negligible for heavy-quark decays. However, its contribution might
become more relevant for kaon decays, as it will be highligthed in
Section 3.

\indent Assuming a constant vertex function $\Lambda$, the quark
triangle diagram, depicted in Fig. 1, yields (cf. ref. [9])
 \be
    \label{13}
    & & <P_2| \bar{Q}_2 \gamma_{\mu} Q_1 |P_1> = i N_c \Lambda_1 
    \Lambda_2 ~ \int \frac{d^4k}{(2\pi)^4} \cdot \\ \nonumber
    & & Sp \left[ \frac{-\hat{k} + m}{k^2 - m^2 + i\varepsilon} \gamma_5
    \frac{\hat{P}_2 - \hat{k} + m_2}{(P_2 - k)^2 - m_2^2 + i\varepsilon}
    \gamma_{\mu} \frac{\hat{P}_1 - \hat{k} + m_1}{(P_1 - k)^2 - m_1^2 +
    i\varepsilon} \gamma_5 \right]
 \ee
where $p_1 \equiv P_1 - k$ and $p_2 \equiv P_2 - k$ are the
four-momenta of the active quarks, with $k$ being the four-momentum of
the spectator quark. In eq. (\ref{13}) $m_1$ and $m_2$ are the masses of
the active quarks, whereas $m$ is the mass of the spectator quark;
finally, $N_c$ is the number of colours. After changing the variables
from $(k_0, {\bf k})$ to $(k^-, k^+, {\bf k_{\bot}})$ with $k^{\pm}
\equiv k_0 \pm k_3$, one obtains
 \be
    \label{14}
    & & <P_2| \bar{Q}_2 \gamma_{\mu} Q_1 |P_1> = \frac{i N_c \Lambda_1 
    \Lambda_2}{2 (2 \pi)^4} \int \frac{dk^- dk^+ d^2k_{\bot}}{k^+ (k^+ -
    P_2^+)(k^+ - P_1^+)} ~ I_{\mu} \cdot \\ \nonumber
    & & \left \{ (k^- - \frac{m_{\bot}^2 - i\varepsilon}{k^+}) (k^- -
    P_2^- - \frac{m_{2\bot}^2 - i\varepsilon}{k^+ - P_2^+}) (k^- - P_1^-
    - \frac{m_{1\bot}^2 - i\varepsilon}{k^+ - P_1^+}) \right \}^{-1}
 \ee
where $m_{i\bot}^2 = k_{\bot}^2 + m_i^2$ ($i=1, 2$), $m_{\bot}^2 =
k_{\bot}^2 + m^2$ and $I_{\mu} = Sp \{ (\hat{k} + m) (\hat{P}_2 -
\hat{k} + m_2) \gamma_{\mu} (\hat{P}_1 - \hat{k} + m_1) \}$. For sake
of simplicity, in eq. (\ref{14}) and in what follows, ${\bf P_{1 \perp}}
= {\bf P_{2 \perp}} = 0$ is assumed.

\indent Let us now calculate eq. (14) for the plus component of the weak
vector current, because for this component the $k^-$-integration is
convergent. Applying the Cauchy theorem, four different cases should be
analyzed: $k^+ < 0$, $0 < k^+ <P_2^+$, $P_2^+ < k^+ <P_1^+$ and $k^+ >
P_1^+$. The first and fourth cases do not contribute to the integral
over $k^-$, because the three poles in eq. (\ref{14}) have imaginary
parts with the same sign. It can be easily seen that the only surviving
contributions come from the regions $0 < k^+ < P_2^+$ (i.e., $0 < x <
y$) and $P_2^+ < k^+ < P_1^+$ (i.e., $y < x <1$), where $x \equiv k^+ /
P_1^+$ is the LF momentum fraction of the spectator quark in the parent
meson. The first region corresponds to the so-called spectator pole,
i.e. to the situation in which the spectator quark is on its mass
shell. Therefore, its plus component of the momentum cannot exceed that
of $P_2$ (i.e., $x / y < 1$). The second region corresponds to the
final active quark on its mass shell. We will refer to these two
contributions as the partonic and the non-partonic ones
\footnote{One could alternatively carry out the integration in eq.
(\ref{13}) over the variable $k^0$, obtaining six different
(time-ordered) diagrams, which are the instant-form representation of
the triangle diagram [13].}. Hereafter, the partonic and the
non-partonic terms will be denoted as $J_A^+$ and $J_B^+$,
respectively. Performing all the traces, one obtains
 \be
    \label{15}
    J_A^+ & = & \frac{N_c \Lambda_1 \Lambda_2}{2 (2\pi)^3} \int_0^y 
    \frac{dx}{x (1 - x) (1 - x')} \int d^2k_{\bot} \frac{I_A^+}{(M_1^2 
    - M_{10}^2) (M_2^2 - M_{20}^2)} \\ \label{16}
    J_B^+ & = & \frac{N_c \Lambda_1 \Lambda_2}{2 (2\pi)^3} \int_y^1 
    \frac{dx}{x (1 - x) (1 - x')} \int d^2k_{\bot} \frac{I_B^+}{(M_1^2 -
    M_{10}^2) (M_{12}^2 - q^2) \frac{y}{1-y}}
 \ee
with $x' \equiv x/y$ and
 \be
    \label{17}
    M_{10}^2 & \equiv & M_{10}^2(x, k^2_{\bot}) = \frac{m^2 +
    k^2_{\bot}}{x} + \frac{m^2_1 + k^2_{\bot}}{1-x} \\ \label{18}
    M_{20}^2 & \equiv & M_{20}^2(x', k^2_{\bot}) = \frac{m^2 +
    k^2_{\bot}}{x'} + \frac{m_2^2 + k^2_{\bot}}{1-x'} \\ \label{19}
    M_{12}^2 & \equiv & M_{12}^2(x, x', k^2_{\bot}) = \left [
    \frac{m^2_1 + k^2_{\bot}}{1 - x} - \frac{m^2_2 + k^2_{\bot}}{y-x}
    \right ] (1 - y)  \\ \label{20}
    I_A^+ & \equiv & I^+|_{k^- = \frac{m^2 + k^2_{\bot}}{x P_1^+}} = 4m
    m_1 m_2 \left[ (1 + v \cdot v_2) v_1^+ + (1 + v \cdot v_1) v_2^+ +
    (1 - v_1 \cdot v_2) v^+ \right] \\ \label{21}
    I_B^+ & \equiv & I^+|_{k^- = P_1^- - \frac{m_1^2 + k^2_{\bot}}{(1 -
    x) P_1^+}} = 4M_1 m_1 m_2 \left[ (U_1 \cdot v_2 - \xi) v_1^+ +
    (U_1 \cdot v_1 - \xi) v_2^+ + \right. \\ \nonumber
    & & \left . (1 - v_1 \cdot v_2) U_1^+ \right]
 \ee
where $\xi \equiv (m_1 - m) / M_1$, $v_i \equiv p_i / m_i$ are
the four-velocities of the active quarks and $v \equiv k / m$ is the
four-velocity of the spectator quark.

\indent One can easily recognize that the expressions $J_A^+$ (eq.
(\ref{15})) and $J_B^+$ (eq. (\ref{16})) are the contributions of the LF
diagrams (A) and (B), depicted in Fig. 1, respectively. It should be
stressed that the term $J_B^+$ corresponds to the creation of a virtual
$\bar{Q}_2 Q_2 $ pair from the vacuum and to the subsequent conversion
of the $\bar{Q}_2 Q_1$ pair into a $W$-boson. Because of the integration
limits in eq. (\ref{16}), $J_B^+$ is relevant only for kinematics
corresponding to a non-vanishing longitudinal momentum transfer
(i.e., for $q^+ \neq 0$ which means $y < 1$), and it does not contribute
when the momentum transfer is purely transverse (i.e., for $q^+ = 0$
which implies $y = 1$).

\indent For a finite range vertex the following substitutions should be
performed in eqs. (\ref{15}-\ref{16}) (cf. ref. [9])
 \be
    \label{22}
    \frac{1}{(2\pi)^{3/2}} ~ \frac{\sqrt{N_c} ~ \Lambda_1}{(1 - x)
    (M_1^2 - M_{10}^2)} \to \chi_1(x, k^2_{\bot}) \\ \label{23}
    \frac{1}{(2\pi)^{3/2}} ~ \frac{\sqrt{N_c} ~ \Lambda_2}{(1 - x')
    (M_2^2 - M_{20}^2)} \to \chi_2(x', k^2_{\bot})
 \ee
where the normalization of the functions $\chi_i$ will be fixed later on
(see eq. (\ref{27})). It can be easily checked that $J_A^+$ involves
the wave functions $\chi_i$ of both the initial and final mesons,
whereas the wave function $\chi_2$ of the final meson cannot be
reconstructed in $J_B^+$, because in the non-partonic diagram the LF
fraction $x / y$ of the spectator quark in the final state would exceed
$1$.

\subsection{The Hamiltonian light-front approach}

Within the LF formalism the parent pseudoscalar meson $Q_1 \bar{q}$ is
represented by the following state vector
 \be
    \label{24}
    |P_1> = \int \frac{d^2k_{\bot}}{\sqrt{16\pi^3}} \cdot \frac{dx}{x}
    [M^2_{10} - (m_1 - m)^2]^{1/2} ~ R_{00}^{(1)}(x, {\bf k_{\bot}}) ~
    \chi_1(x, k^2_{\bot}) ~ Q_1^{\dagger} \bar{q}^{\dagger} |0>
 \ee
where $Q_1^{\dagger} = Q_1^{\dagger}(\vec{p}_1, \lambda)$ and
$\bar{q}^{\dagger} = \bar{q}^{\dagger}(\vec{k}, \bar{\lambda})$ are
the creation operators for a heavy quark and a spectator (anti-) quark
with the following anticommutation rules
 \be
    \label{25}
    \{ Q_1^{\dagger}(\vec{p}_1, \lambda), Q_1(\vec{p'}_1, \lambda')
    \} = (2\pi)^3 ~ 2p_1^+ ~ \delta_{\lambda \lambda'} ~
    \delta(\vec{p}_1 - \vec{p'}_1) \\ \nonumber
    \{ \bar{q}^{\dagger}(\vec{k}, \bar{\lambda}), ~
    \bar{q}(\vec{k'}, \bar{\lambda'}) \} = (2\pi)^3 ~ 2k^+ ~
    \delta_{\bar{\lambda} \bar{\lambda'}} ~ \delta(\vec{k} - \vec{k'})
 \ee
where $\vec{k} = (k^+, {\bf k_{\bot}})$ and $\vec{p}_1 \equiv \vec{P}_1
- \vec{k} = (P_1^+ - k^+, {-\bf k_{\bot}})$ are LF momenta. The
spectator quark carries the fraction $x$ of the plus component of the
meson momentum, while the heavy quark carries the fraction $1 - x$. In
eq. (\ref{24}) $R_{00}^{(1)}(x, {\bf k_{\bot}})$ is a momentum-dependent
quantity arising from the Melosh rotations of the quark spins (see
below eq. (\ref{31})), and $M_{10}^2$ has been already defined in eq.
(\ref{17}). The state vector $|\vec{P}_2>$ of the daughter meson $Q_2
\bar{q}$ has a form analogous to eq. (\ref{24}) with the obvious
replacements $m_1 \to m_2 $, $M^2_{10} \to M^2_{20}$, $R_{00}^{(1)} \to
R_{00}^{(2)}$, $\chi_1 \to \chi_2$ and $Q_1^{\dagger} \to
Q_2^{\dagger}$. The state vectors $|P_i>$ ($i=1, 2$) are normalized as
 \be
    \label{26}
    <P'_i|P_i> ~ = (2\pi)^3 ~ 2P_i^+ ~ \delta(\vec{P}_i - \vec{P'}_i)
 \ee
so that the normalization of the wave functions $\chi_i(x, k^2_{\bot})$
reads as
 \be
    \label{27}
    \int_0^1 dx \frac{1 - x}{x} \int d^2k_{\bot} ~ [M^2_{i0} - (m_i -
    m)^2] ~ \chi_i^2(x, k^2_{\bot}) = 1
 \ee
Following the Brodsky-Huang-Lepage prescription [12] the function
$\chi_i(x, k^2_{\bot})$ can be related to the equal-time wave function
$w_i(k^2)$ normalized according to
 \be
    \label{28}
    \int_0^{\infty} dk ~ k^2 ~ w_i^2(k^2) = 1
\ee
provided that the fraction $x$ is replaced by the longitudinal momentum
$ k^{(i)}_3 $ defined as
 \be
    \label{29}
    k^{(i)}_3 = \left( x - {1 \over 2} \right) M_{i0} + \frac{m_i^2 -
    m^2}{2M_{i0}}
 \ee
Explicitly, one has
 \be
    \label{30}
    \chi_i(x, k^2_{\bot}) = \frac{1}{2 (1 - x)} \frac{\sqrt{M_{i0} [1 -
    (m_i^2 - m^2)^2 / M^4_{i0}]}} {\sqrt{M^2_{i0} - (m_i - m)^2}} ~
    \frac{w_i(k^2)}{\sqrt{4\pi}}
 \ee
with $k^2 \equiv k_{\perp}^2 + (k^{(i)}_3)^2$. Finally, using the chiral
representation of the Dirac spinors (see, e.g., ref. [9]), one has
 \be
    \label{31}
    [R_{00}^{(i)}(x, {\bf k_{\bot}})]_{\lambda \bar{\lambda}} =
    \frac{1}{\sqrt{2} ~ \sqrt{M_{i0}^2 - (m_i - m)^2}} ~
    \bar{u}(\vec{p}_i, \lambda) \gamma_5 v(\vec{k}, \bar{\lambda})
 \ee
Thus, the matrix element of the "good" current $J^+ = J_0 + J_3$ reads
as 
 \be
    \label{32}
    J^+(y) = \int_0^y \frac{dx}{2x} \int d^2k_{\bot} ~ \chi_1(x,
    k^2_{\bot}) ~ \chi_2 \left( {x \over y}, k^2_{\bot} \right) ~
    I_{LF}^+(x, y, k^2_{\bot})
 \ee
where $y = y_1$ or $y = y_2$ (see eq. (3)) and $I_{LF}^+$ is the spin
contribution
 \be
    \label{33}
    I_{LF}^+ & = & Sp \left \{ (-\hat{k} + m) \gamma_5 (\hat{P}_2 -
    \hat{k} + m_2) \gamma_+ (\hat{P}_1 - \hat{k} + m_1) \gamma_5 \right
    \} \\ \label{34}
    & = & 4m m_1 m_2 \left \{ (1 + v \cdot v_2) v_1^+ + (1 + v_1 \cdot
    v) v_2^+ + (1 - v_1 \cdot v_2) v^+ \right \}
 \ee
It can be seen that eq. (\ref{34}) coincides with eq. (\ref{20}).
Moreover, it can be easily checked that, using eqs.
(\ref{22}-\ref{23}) the partonic term $J_A^+$ coincides with eq.
(\ref{32}). In other words, the result (\ref{32}), obtained within
the Hamiltonian LF dynamics, is equivalent to the spectator pole term of
the quark triangle diagram. This result generalizes to the time-like
region ($q^2 > 0$) the result obtained in ref. [11] in the Breit frame
($q^2 \leq 0$). In terms of the LF variables $x$ and $k^2_{\bot}$ one
gets
 \be
    \label{35}
    I_{LF}^+ = \frac{4 P_1^+}{x'} \{ [m (1 - x) + x m_1] [m (1 - x') +
    x' m_2] + k_{\bot}^2 \}
 \ee
The form factors $h_{\pm}$, or $f_{\pm}$, can be evaluated using 
only the matrix elements of the "good" component of the current $J^+ =
\bar{Q}_2 \gamma_+ Q_1$. In order to invert eq. (\ref{6}) we calculate
the matrix element of $J^+$ in two reference frames having the 3-axis
parallel and anti-parallel to the 3-momentum of the daughter meson. In
this way two different matrix elements, denoted hereafter as
$J_1^+(q^2) \equiv 2 P_1^+ ~ H_1(q^2)$ and $J_2^+(q^2) \equiv 2 P_1^+ ~
H_2(q^2)$, are obtained. Then, $h_{\pm}$ and $f_{\pm}$ are given by
 \be
    \label{36}
    h_+(q^2) & = & \frac{(\zeta - y_2) H_1(q^2) - (\zeta - y_1)
    H_2(q^2)} {\sqrt{\zeta} ~ (y_1 - y_2)} \\ \label{37}
    h_-(q^2) & = & \frac{(\zeta + y_2) H_1(q^2) - (\zeta + y_1) 
    H_2(q^2)} {\sqrt{\zeta} ~ (y_2 - y_1)} \\ \label{38}
    f_+(q^2) & = & \frac{(1 - y_2) H_1(q^2) - (1 - y_1) H_2(q^2)}{y_1
    - y_2}
    \\ \label{39}
    f_-(q^2) & = & \frac{(1 + y_2) H_1(q^2) - (1 + y_1) H_2(q^2)}{y_2
    - y_1}
 \ee
Note that at the point of zero recoil, i.e. $q^2 = q^2_{max}$, where
$y_1 = y_2 = \zeta$, the two reference frames coincide, so that
$H_1(q^2_{max}) = H_2(q^2_{max})$. This implies that the form factors
have no singularity at $q^2 = q^2_{\max}$. At the point of maximum
recoil, i.e. $q^2 = 0$, the expression of $f_+(0) = H_1(0)$ simplifies
to
 \be
    \label{40}
    f_+(0) & = & \int_0^1 dx \int d^2k_{\bot}
    \sqrt{\frac{dk^{(1)}_3}{dx}} \sqrt{\frac{dk^{(2)}_3}{dx}}
    \frac{w_1(k^2) ~ w_2(k^2)}{4\pi} \cdot \\ \nonumber
    & & \frac{[m (1 - x) + x m_1] [m (1 - x) + x m_2] + k^2_{\bot}}
    {\sqrt{x (1 - x) [M_{10}^2(x) - (m_1 - m)^2]} ~ \sqrt{x (1 - x)
    [M_{20}^2(x) - (m_2 - m)^2]}}
 \ee
where
 \be
    \label{41}
    \frac{dk^{(i)}_3}{dx} = \frac{M_{i0}}{4x (1 - x)} \left[ 1 -
    \left( \frac{m_i^2 - m^2}{M^2_{i0}} \right)^2 \right]
 \ee
Eq. (\ref{40}) can be rewritten as
 \be
    \label{42}
    f_+(0) = \int_0^1 dx \int d^2k_{\bot} \Phi_1(x, k_{\bot}^2) ~
    \Phi_2(x, k_{\bot}^2) \frac{A_1 A_2 + k_{\bot}^2}{\sqrt{(A^2_1 +
    k^2_{\bot}) (A_2^2 + k_{\bot}^2)}}
 \ee
where $A_i \equiv x m_i + (1 - x) m$ and $\Phi_i(x, k^2_{\bot}) =
\sqrt{\frac{dk^{(i)}_3}{dx}} \frac{w_i(k^2)}{\sqrt{4\pi}}$. It can be
seen that eq. (\ref{42}) coincides with the result of refs. [9] and
[15] obtained using the Hamiltonian light-front formalism in the Breit
frame.

\indent Note also that in the non-relativistic limit ($k^2 \ll m^2, ~
m^2_i$ and $k'^2 \ll m^2, m^2_i$) eq. (\ref{32}) reduces to
 \be
    \label{43}
    J^+_{1,2}(q^2) = 2 ~ \sqrt{M_1 M_2} ~ \int d^3k ~ w_1(k^2) ~
    w_2(k^{'2})(1 \mp \frac{k_3}{2m_1} \pm \frac{k_3 + p}{2m_2})
 \ee
where $k^{'2} = k_{\perp}^2 + k_3^{'2}$ and $k'_3 = k_3 - \frac{m}{M_2}
~ p$, with $p$ being the 3-momentum of the daughter meson in the rest
frame of the parent meson. In the non-relativistic limit one has $p =
M_2 (y_1 / \zeta - 1)$. Eq. (\ref{43}) written for the "currents"
$J_0 = \frac{1}{2}(J_1^+ + J_2^+)$ and $J_3 = \frac{1}{2}(J_1^+ -
J_2^+)$  coincide with the non-relativistic result derived in the ISGW
model [7].

\section{Results}

The calculation of the semileptonic form factors $f_{\pm}(q^2)$ (eqs.
(\ref{38}-\ref{39})) has been performed using our LF result (eqs.
(\ref{32}) and (\ref{35})) for computing the matrix elements
$J_{1,2}^+$ and eq. (\ref{30}) for the meson wave functions $\chi_i$. As
for the radial wave function $w_i(k^2)$ appearing in eq. (\ref{30}),
the Gaussian ansatz of the ISGW model has been adopted; the values of
the parameters (the constituent quark masses and the harmonic
oscillator (HO) lengths) are taken from the updated version of this
model [16] \footnote{In what follows this version will be referred
to as the ISGW2 model.}.

\indent The $q^2$ behaviour of the form factors $f_{\pm}$ for the $b \to
c \ell \nu_{\ell}$, $b \to u \ell \nu_{\ell}$, $c \to s \ell
\nu_{\ell}$, $c \to u \ell \nu_{\ell}$ and $s \to u \ell \nu_{\ell}$
quark decays are shown in Figs. 2-6, respectively. The solid lines are
the results of our LF calculations obtained using the ISGW2 values for
the quark masses and HO parameters, but adopting the physical values
for the meson masses taken from PDG '94 [17]. For comparison, the
results obtained using all the ISGW2 parameters, i.e. adopting also the
meson masses used in [16], are shown by the dashed lines. The dotted
lines are the monopole approximations for the form factors (eq.
(\ref{11})), obtained using the value of the form factors at $q^2 = 0$
and the pole masses given by the lowest-lying vector meson mass for the
given channel [6]. It can be seen that: i) our simple pole approximation
yields a $q^2$-dependence which sharply differs from the one obtained
within our LF approach, particularly in case of heavy-to-light decays,
like $B \to \pi$ and $D \to \pi$; for these decays the form factors do
not obey at all the pattern of pole dominance near the zero-recoil
point ($\eta = 1$); ii) for the heavy-to-light decays $B \to \pi$ and $D
\to \pi$, as well as for the $K_{e3}$ decay, the form factors are
strongly sensitive to the choice of the values of the meson masses
(compare dashed and solid lines); it is worth nothing that this happens
not only near the kinematical end-point of maximum recoil, but in the
whole accessible kinematical region.

\indent The results for the semileptonic decay rate $\Gamma$ (eq.
(\ref{12})) are collected in Table 1. It can be seen that: i) the
monopole approximation of ref. [6] underestimates the rates, ranging
from $\sim 5\%$ for $K \to \pi$ to $\sim 20\%$ for $B \to D$ and to
$\sim 40\%$ for $B \to \pi$ ; ii) the choice of the values of the meson
masses is crucial for the $K_{e3}$ decay.

\indent We now present our results, organized by the underlying quark
decay and arranged in order of decreasing active quark mass. We will
compare our results to experimental data and to predictions of
different models.

\subsection{Decay $B \to D \ell \nu_{\ell}$}

The $q^2$ behaviour of the form factor $f_{\pm}$ for the semileptonic
$B \to D$ transition is shown in Fig. 2. The slope $\hat{\rho}^2$ of
the form factor $h_+$ (eq. (\ref{36})) at the point of zero recoil,
defined as $h_+(\eta) \approx h_+(1) ~ (1 - \hat{\rho}^2 (\eta - 1))$,
turns out to be $\hat{\rho}^2 = 1.3$ within our LF approach. As is
known, the slope parameter $\hat{\rho}^2$ of the physical form factor
$h_+$ differs from the slope parameter $\rho^2$ of the universal
Isgur-Wise function by corrections that violate the heavy-quark
symmetry. Using an approximate relation between the two quantities,
namely $\hat{\rho}^2 \approx \rho^2 + 0.2$ [18], we obtain $\rho^2 =
1.1$. This value compares well with the quark-model prediction of ref.
[19] and it is only slightly higher than QCD sum rule predictions,
which typically range from $0.7$ to $1.0$ (see, e.g., ref. [20]).
Performing the calculations with and without the effects of the Melosh
composition of quark spins (i.e., assuming $R_{00}^{(1)} \neq 1$ or
$R_{00}^{(1)} =1$ in eq. (\ref{24})), it turns out that the effects of
the Melosh rotations increase $\rho^2$ by $\sim  20\%$; this result
agrees with the conclusion of ref. [19] obtained in the zero-binding
approximation.

\indent From the measured rate \footnote{This value has been derived
by combining the branching ratio $Br(B^0 \to D^- \ell^+ \nu_ell)  =
(1.9 \pm 0.5) \%$ with the world average of the $B^0$ lifetime
$\tau_{B^0} = 1.50 \pm 0.11 ~ ps$ [17].}
 \be
    \label{3.1}
    \Gamma (B^0 \to D^- \ell^+ \nu_{\ell}) = (1.27 \pm 0.43) \cdot
    10^{10} ~ s^{-1}
 \ee
and our predicted rate from Table 1, we obtain
 \be
    \label{3.2}
     |V_{bc}| = 0.036 \pm 0.004
\ee
Our LF prediction is only $\sim 10 \%$ larger than the corresponding
ISGW2 prediction ($|V_{bc}| = 0.033 \pm 0.004$ [16]) and in agreement
with the updated "experimental" determinations of $|V_{bc}|$ [21]
obtained from exclusive and inclusive semileptonic decays of the B-meson
($|V_{bc}|_{excl} = 0.0373 \pm 0.0045_{exp} \pm 0.0065_{th}$ and 
$|V_{bc}|_{incl} = 0.0398 \pm 0.0008_{exp} \pm 0.0040_{th}$).

\indent Within an accuracy of few percent our form factor $f_+(q^2)$ is
reproduced by a monopole approximation with $M_{pole} = 4.82 ~ GeV$. As
a matter of fact, at a level of accuracy of $\sim 10 \%$ for the decay
rate, the details of the $q^2$ dependence of $f_+$ do not matter as
long as $q^2 \ll M_{pole}^2$ and the form factor varies slowly with
$q^2$, as it is the case for the semileptonic $B \to D$ decay.

\subsection{Decay $B \to \pi \ell \nu_{\ell}$}

We now consider the semileptonic B-meson decay corresponding to the
quark process $b \to u \ell \nu_{\ell}$. The investigation of this
decay is very important for the determination of the $|V_{bu}|$ CKM
matrix element, which plays an important role for the CP violation in
the Standard Model. The experimental studies of such a decay have
begun and, very recently, the CLEO Collaboration [22] has reported the
first signal for exclusive semileptonic decays of the B meson into
charmless final states, in particular for the decay mode $B \to \pi
\ell \nu_{\ell}$. However, there is a significant model dependence in
the simulation of the reconstruction efficiencies. The observed
branching ratios,  extracted adopting the ISGW [8] and WSB [6] models,
is
  \be
     \label{3.3}
     Br(B \to \pi \ell \nu_{\ell}) & = & (1.34 \pm 0.45) \cdot 10^{-4}
     ~~~~~~~~ ISGW \\ \label{3.4}
     Br(B \to \pi \ell \nu_{\ell}) & = & (1.63 \pm 0.57) \cdot 10^{-4}
     ~~~~~~~~ WSB
 \ee
Combining the average of these results with the world average value of
the $B^0$ lifetime, one gets
 \be
    \label{3.5}
    \Gamma (B^0 \to \pi^- \ell^+ \nu_{\ell}) = (0.99 \pm 0.25) \cdot
    10^{-4} ~ ps^{-1}
 \ee
From our predicted rate $\Gamma (B^0 \to \pi^- \ell^+ \nu_{\ell}) =
9.62 ~ |V_{bu}|^2 ~ ps^{-1}$ (see Table 1), we obtain
 \be
    \label{3.6}
    |V_{bu}| = 0.0032 \pm 0.0004
 \ee
Using our previous result (\ref{3.2}) for $|V_{bc}|$ we get
 \be
     \label{3.7}
    \left| \frac{V_{bu}}{V_{cb}} \right| = 0.088 \pm 0.015
\ee
which is in nice agreement with the value derived from measurements of
the end-point region of the lepton spectrum in inclusive semileptonic
decays [23, 24], viz.
 \be
    \label{3.8}
    \left| \frac{V_{bu}}{V_{cb}} \right|_{incl} = 0.08 \pm 0.01_{exp}
    \pm 0.02_{th}
 \ee
A large variety of calculations of the ratio $|V_{bu} / V_{cb}|$, based
on various non-perturbative approaches, exists in the literature; the
results typically lie in the range from $0.06$ to $0.11$.

\indent The $q^2$ behaviour of the form factors $f_{\pm}$ for the
semileptonic $B \to \pi$ transition is shown in Fig. 3. Our result for $
f^{B \to \pi}_+(q^2 = 0)$ (see Table 2) nicely agrees with the one
obtained from a recent analysis of the $B \to \pi$ form factors using
the light-front QCD sum rule [5]. Other model predictions for $f^{B \to
\pi}_+(0)$ and the $B \to \pi \ell \nu_{\ell}$ decay rate are collected
in Tables 2 and 3.

\subsection{Decay $D \to K \ell \nu_{\ell}$}

The value of $|V_{cs}|$ can be extracted from measurements of charmed
hadron production in neutrino experiments. However, such a procedure
depends crucially on the assumption about the strange quark density in
the partonic sea. The most conservative assumption (i.e., an $SU(3)$
symmetric sea) leads to the bound $|V_{cs}| > 0.59 $. Therefore, it is
better to proceed in a way analogous to the method used for extracting
$|V_{bc}|$ and $|V_{bu}|$ from $B$-meson decays. By combining the
experimental data on the branching ratios for the semileptonic
$D^0$ decay ($Br(D^0 \to K^- e^+ \nu_e) = 3.68 \pm 0.21 \%$ [17])
with the accurate value of the $D^0$ lifetime ($\tau^{D^0} = 0.415 \pm
0.004 ~ ps$ [17]), one has
 \be
    \label{3.10}
    \Gamma (D^0 \to K^- e^+ \nu_e) = (0.089 \pm 0.005) ~ ps^{-1}
 \ee
From our predicted rate $\Gamma (D^0 \to K^- e^+ \nu_e) = 0.113 \cdot
|V_{cs}|^2  ~ ps^{-1}$ it follows 
 \be
    \label{3.11}
    |V_{cs}| = 0.89 \pm 0.03
 \ee
The constraint of unitarity of the CKM matrix with three generations of
leptons gives a much higher value, viz. $|V_{cs}| = 0.974$ [17].

\indent The $q^2$ behaviour of the form factors $f_{\pm}$ for the $D \to
K$ transition is shown in Fig. 4. At $q^2 = 0$ we have obtained: $f^{D
\to K}_+(0) = 0.78$ and $f^{D \to K}_+(q^2_{max}) = 1.56$ (see Table 1).
Our value for $f^{D \to K}_+(0)$ compares favourably with the
experimental average $0.75 \pm 0.03$ [34], as well as with the recent
experimental results [35] $f^{D \to K}_+(0) = 0.77 \pm 0.01 \pm 0.04$
and $f^{D \to K}_+(q^2_{max}) = 1.42 \pm 0.25$ \footnote{For this
decay CLEO [35] has obtained for the pole mass of the $f_+$ form factor
the value $M_{pole} = 2.00 \pm 0.12 \pm 0.18 ~ GeV$, assuming a
monopole shape. Using the measured lepton spectrum, the decay rate can
be transformed into a measurement of $f_+(q^2 = 0)$.}. For comparison,
the ISGW2 predictions are $f^{D \to K}_+(0) = 0.85$ and $f^{D \to
K}_+(q^2_{max}) = 1.23$, respectively [8]. Very recently, the E687
collaboration has investigated the $D^0 \to K^- \mu^+ \nu_{\mu}$ and
$D^0 \to K^- \pi^+$ decays and has reported [36] the following values:
$f^{D \to K}_+(0) = 0.71 \pm 0.03 \pm 0.03$ and $f^{D \to K}_-(0) /
f^{D \to K}_+(0) = -1.3_{-0.34}^{+0.36} \pm 0.6$. The latter result is
reproduced by our calculations, yielding the value $-1.06$. It is worth
noting that the lattice QCD simulations of ref. [32] gives the value
$-1.2 \pm 0.5$, whereas both the WSB [6] and ISGW [7] quark model
predictions are much lower, namely $-0.46$ and $-0.60$, respectively.

\subsection{Decay $D \to \pi \ell \nu_{\ell}$}

This is the only heavy-to-light decay where a comparison with experiment
is possible. Moreover, there exist several model calculations in the
literature, using QCD sum rules [3-5, 30], quark models [6-9,16], and
few lattice QCD calculations [31, 32]. The results of ref. [29] are
based on the HQET, adopting the experimental results [33] as input for
the values of the form factors of the $B \to \pi e \nu_e$ and $B \to
\rho e \nu_e$ transitions. Our predictions for the Cabibbo suppressed
decay $D \to \pi$ are shown in Fig. 5 and Table 1. Our form factor
$f_+(q^2)$ turns out to be flat near $q^2 = q_{max}^2$ with a negative
slope. Assuming $|V_{cd}| = 0.221 \pm 0.003$ [17], which is inferred
from the unitarity of the CKM matrix, we predict for the semileptonic
decay rate the following value
 \be
    \label{3.12}
     \Gamma (D^0 \to \pi^- e^+ \nu_e) = 7.8 \cdot 10^{-3} ~ ps^{-1}
 \ee
This prediction should be compared with the experimental result
\be
   \label{3.13}
   \Gamma_{exp} (D^0 \to \pi^- e^+ \nu_e) = (9.4^{+5.5}_{-2.9}) \cdot
   10^{-3} ~ ps^{-1}
 \ee
and the recent light-front QCD sum rule prediction [5]
\be
   \label{3.14}
   \Gamma_{SR} (D^0 \to \pi^- e^+ \nu_e) = (7.6 \pm 0.2) \cdot 10^{-3} 
   ~ ps^{-1}
 \ee
The comparison of our values for $f_+^{D \to \pi}(0)$ and the decay
rate $\Gamma$ with other theoretical results is presented in Tables 2
and 3.

\indent The ratio of the branching ratio of the Cabibbo suppressed decay
$D \to \pi \ell \nu_{\ell}$ to that of the Cabibbo favoured decay $D \to
K \ell \nu_{\ell}$ is found to be $1.426 ~ |\frac{V_{cd}}{V_{cs}}|^2$.
Using the unitarity constraint on the CKM matrix, yielding
$|\frac{V_{cd}}{V_{cs}}|^2 = 0.051 \pm 0.001 $, we obtain the value 
 \be
    \label{3.15}
    R_0 = \frac{Br(D \to \pi \ell \nu_{\ell})}{Br(D \to K \ell
    \nu_{\ell})} = 7.3 \% 
\ee
which is slightly lower than the predictions from various theoretical
models, namely the CQM $(R_0 = 8.8 \%$ [6], $R_0 = 5.3 \%$ [8]), the
QCD sum rules $(R_0 = 9.3 \%$  [30], $R_0 = 8.3 \%$ [37]), and lattice
QCD calculations $(R_0 = 8.6 \%$ [31]).

\indent The $D \to K \ell \nu_{\ell} $ and $D \to \pi \ell \nu_{\ell}$
decays have been recently investigated by the CLEO-II collaboration and
both charged [38] and neutral [39] $D$-meson decays have been measured.
The results are
 \be
    \label{3.16}
    \frac{Br(D^+ \to \pi^0 e^+ \nu_e)}{Br(D^+ \to \bar{K}^0 e^+ \nu_e)}
    = (8.5 \pm 2.7 \pm 1.4) \%   ~~~~~~~~ [38] \\ \label{3.17}
    \frac{Br(D^0 \to \pi^- e^+ \nu_e)}{Br(D^0 \to K^- e^+ \nu_e)} =
    (10.3 \pm 3.9 \pm 1.3) \% ~~~~~~~~ [39]
 \ee
Assuming isospin invariance (i.e., $Br(D^+ \to \pi^0 e^+ \nu_e) /
Br(D^0 \to \pi^- e^+ \nu_e) = 1 /2$) and the pole dominance for the
$q^2$-dependence of the form factors, with the mass of the vector
resonance given by the mass of the $D^*$ ($D^*_s$) meson for the $\pi e
\nu_e$ ($K e \nu_e$) decay, these results can be translated into the
following values 
 \be
    \label{3.18}
    \left| \frac{f^{D \to \pi}_+(0)}{f^{D \to K}_+(0)} \right| & = &
    1.29 \pm 0.21 \pm 0.11 ~~~~~~~~ [38] \\ \label{3.19}
    \left| \frac{f^{D \to \pi}_+(0)}{f^{D \to K}_+(0)} \right| & = &
    1.01 \pm 0.20 \pm 0.07 ~~~~~~~~ [39]
 \ee
With respect to the average of these results we predict a slightly lower
ratio, viz.
 \be
    \label{3.20}
    \left| \frac{f^{D \to \pi}_+(0)}{f^{D \to K}_+(0)} \right| = 0.87
\ee
The ISGW2 value for this ratio is $0.71$ and other model predictions
typically range from $0.7$ to $1.4$.

\subsection{$ K \to \pi e \nu_e $}

The non-partonic contribution $J_B^+$ to the matrix element of the weak
vector current might be more important for light meson decays than for
the heavy-to-heavy and heavy-to-light transitions. However, the
corrections arising from the non-partonic diagram are expected to be
relevant mainly for the form factor $f_-(q^2)$. Therefore, it is of
interest to consider the $K_{e3}$ decay, whose differential decay rate
is governed by $f_+(q^2)$ only. The form factors for $K_{e3}$ decays
are usually referred at the $SU(3)$ normalization point $q^2 = 0$. Our
result $f_+(0) = 0.976$ is in nice agreement with the "standard" value
$f_+(0) = 0.97 \pm 0.01$ [40]. Our result for $ f_+(q^2_{max}) = 1.120
$ is roughly consistent with the Ademollo-Gatto theorem [41], which
protects $f_+(q^2_{max})$ from substantial deviations from unity. The
$q^2$ behaviour of the form factor $f_+$ for the $ K \to \pi $
transitions is shown in Fig. 6. Using $|V_{us}| = 0.2205 \pm 0.0018$
[17], we obtain $\Gamma(K_{e3}) = (7.58 \pm 0.13) \cdot 10^6 ~ s^{-1}$
in agreement with the experimental average value $(7.7 \pm 0.5) \cdot
10^6 ~ s^{-1}$ [17].

\section{Conclusions}

The weak transition form factors, which govern the heavy-to-heavy,
heavy-to-light and $K_{e3}$ semileptonic decays of pseudoscalar
mesons, have been investigated within a relativistic constituent quark
model based on the light-front formalism. Using the "good" component
of the weak vector current, it has been shown that the partonic term
of the quark triangle diagram is equivalent to the result which can
be obtained within the Hamiltonian light-front dynamics, generalizing
in this way to the time-like region a previous result [11] derived
only for space-like values of the momentum transfer. For the numerical
investigations, the equal-time wave function of the ISGW model has been
adopted, so that, for the first time, the transition form factors have
been calculated in the whole kinematical region accessible in
semileptonic decays. We have calculated the form factors and the decay
rate for the $B \to D \ell \nu_{\ell}$, $B \to \pi \ell \nu_{\ell}$, $D
\to K \ell \nu_{\ell}$, $D \to \pi \ell \nu_{\ell}$ and $K \to \pi \ell
\nu_{\ell}$ weak decays. The relevance of the use of the physical
values of the meson masses, as well as the possible limitations of the
pole dominance approximation, have been illustrated. Our results have
been successfully compared  with available experimental data and
predictions from different approaches. In particular, using the
available experimental information on the semileptonic decay rates, the
CKM parameters have been estimated. With the only exception of
$|V_{cs}|$ our results are in good agreement with existing
determinations of these parameters. Before closing, it should be
reminded that in our calculations the contribution of the pair creation
from the vacuum has been neglected; therefore, an estimate of such
contribution, particularly in case of the heavy-to-light and $K_{\ell
3}$ transitions, is mandatory for a complete comparison with
experimental data.

\section*{Acknowledgement}

Two of the authors (I.L.G. and I.M.N.) acknowledge the financial support
of the INTAS grant No 93-0079. This work was done in part under the RFFR
grant, Ref. No. 95-02-04808a.

\newpage

\centerline{\bf Table 1}

\noindent Form factor $f_+(q^2)$ (eq. (\ref{38})), evaluated at $q^2 =
0$ and $q^2 = q^2_{\max}$, and the decay rate $\Gamma$ (eq. (\ref{12}))
for various semileptonic decays (in units $ps^{-1}$). $\Gamma'$ is the
same as $\Gamma$, but using the meson masses from ref. [16] instead of
their experimental values. $\Gamma_{pole}$ denotes the decay rate
calculated using the pole approximation for $f_+(q^2)$ (eq. (\ref{11}))
with the pole masses taken from [6]. \\

\bc

\begin{tabular}{|l|ll|lll|}\hline
weak       & $f_+(0)$ & $f_+(q^2_{\max})$ & $\Gamma$ & $\Gamma'$ &
$\Gamma_{pole}$ \\ transition &&&&& \\ \hline

  $B \to D$& 0.684 &1.365 & 9.78$|V_{bc}|^2$ & 9.78$|V_{bc}|^2$ &
7.89$|V_{bc}|^2$ \\

$B \to \pi$& 0.293 &1.658 & 9.62$|V_{bu}|^2$ & 9.05$|V_{bu}|^2$ &
5.80$|V_{bu}|^2$ \\

  $D \to K$& 0.780 &1.560 &0.113$|V_{cs}|^2$ &0.094$|V_{cs}|^2$ &
0.095$|V_{cs}|^2$ \\

$D \to \pi$& 0.681 &1.289 &0.160$|V_{cu}|^2$ &0.121$|V_{cu}|^2$ &
0.142$|V_{cu}|^2$ \\

   $K_{e3}$& 0.976 &1.119 & 1.56$\cdot10^{-4}|V_{su}|^2$ &
0.20$\cdot10^{-4}|V_{su}|^2$ & 1.48$\cdot10^{-4}|V_{su}|^2$ \\ \hline
\end{tabular}

\ec

\newpage

\centerline{\bf Table 2}

\noindent The form factor $f_+$ (eq. (\ref{7})) for the $b \to u$ and
$c \to d$ transitions at $ q^2 = 0 $ in different models.\\

\bc

\begin{tabular}{|ll|ll|}\hline
Reference & $f_+^{B\to\pi}(0)$ & Reference & $f^{D\to\pi}_+(0)$\\ \hline
This paper & 0.293 & This paper & 0.684\\

 [5]$^a$ & $0.29\pm0.01$ & [5]$^a$ & $0.66\pm0.03          $\\

 [3]$^a$ & $0.26\pm0.02$ & [3]$^a$ & $0.5 \pm0.1           $\\

[25]$^a$ & $0.26\pm0.01$ & [30]$^a$ & $0.75\pm0.05         $\\

[26]$^a$ & $0.23\pm0.02$ &  [6]$^b$ & $0.69                $\\

[27]$^a$ & $0.4 \pm0.1 $ &  [8]$^b$ & $0.51                $\\

 [6]$^b$ & $0.33       $ & [29]$^c$ & $0.79                $\\

 [8]$^b$ & $0.09       $ & [31]$^d$ & $0.58\pm0.09         $\\

[28]$^b$ & $0.21\pm0.02$ & [32]$^d$ & $0.84\pm0.12\pm0.35  $\\

[29]$^c$ & $0.89       $ & [33]$^e$ & $0.80^{+0.21}_{-0.14}$\\ \hline
\end{tabular}

\ec

\bigskip

$^a$ QCD sum rules.\\

$^b$ Quark model.\\

$^c$ HQET and chiral perturbation theory.\\

$^d$ Lattice QCD calculations.\\

$^e$ Experimental value.

\newpage

\centerline{\bf Table 3}

\noindent The semileptonic decay rate $\Gamma$ (eq. (\ref{12})) for the
$b \to u$ and $c \to d$ transitions in units $|V_{ub}|^2 \cdot
10^{13}s^{-1}$ and $|V_{cd}|^2 \cdot 10^{11}s^{-1}$, respectively.\\

\bc

\begin{tabular}{|ll|ll|}\hline
Reference & $\Gamma(\bar B^0\to\pi^+ e^-\bar{\nu})$ & Reference &
$\Gamma(D^0\to\pi^-e^+\nu)$\\ \hline

This paper$^a$ & 0.962 (0.79) & This paper$^a$ & 1.60 (1.42) \\

 [5]$^b$&  0.81           &  [5]$^b$& 1.56                   \\

 [3]$^b$& $0.51\pm0.11$   &  [3]$^b$& $0.80\pm0.17$          \\

[25]$^b$& $0.68\pm0.23$   & [30]$^b$& $1.66^{+0.23}_{-0.21}$ \\

[26]$^b$& $0.302\pm0.005$ &  [6]$^c$& 1.41                   \\

[27]$^b$& $1.45\pm0.59$   &  [8]$^c$& 0.77                   \\

 [6]$^c$&  0.74           & [31]$^e$& $0.99^{+0.34}_{-0.28}$ \\

 [8]$^c$&  0.21           & [32]$^e$& $2.09^{+2.24}_{-1.44}$ \\

[28]$^c$& $0.31\pm0.06$   & [33]$^f$&
$|0.22 / V_{cd}|^2 \cdot \left( 1.9^{+1.1}_{-0.6} \right)$   \\

[29]$^d$&  5.4            &         &                        \\ \hline
\end{tabular}

\ec

$^a$ In the parentheses the rates obtained assuming the monopole
approximation for the form factor $f_+$ (eq. (\ref{11})) and the pole
masses from ref. [6], are shown.\\

$^b$ QCD sum rules.\\

$^c$ Quark model.\\

$^d$ HQET and chiral perturbation theory.\\

$^e$ Lattice QCD calculations.\\

$^f$ Experimental value.

\newpage

\section*{Figure Captions}

\begin{description}

\item[Fig. 1] The Feynman triangle diagram for the form factor and the
corresponding light-front diagrams.

\item[Fig. 2] The form factors $f_{\pm}(\eta)$ (eq.
(\ref{38}-\ref{39})) for the $B \to D$ transition. The relation
between the kinematical variables $\eta$ and $q^2$ is given by eq.
(\ref{4}). The solid lines are the results of our LF calculations
obtained using the ISGW2 parameters, but adopting the experimental
values for the meson masses. The results obtained with all the ISGW2
parameters, including the meson masses from ref. [16] are shown by the
dashed lines. The dotted lines are the monopole approximation for the
form factors (eq. \ref{11})), calculated using the values of the form
factor at $q^2 = 0$ and the pole masses given in ref. [6].

\item[Fig. 3] The form factors $f_{\pm}(\eta)$ for the $B \to \pi$
transition. The notations are the same as in Fig. 2.

\item[Fig. 4] The form factors $f_{\pm}(\eta)$ for the $D \to K$
transition. The notations are the same as in Fig. 2.

\item[Fig. 5] The form factors $f_{\pm}(\eta)$ for the $D \to \pi$
transition. The notations are the same as in Fig. 2.

\item[Fig. 6] The form factor $f_+(\eta)$ for the $K \to \pi$
transition. The notations are the same as in Fig. 2.

\end{description}

\newpage

\begin{figure}

\epsfig{file=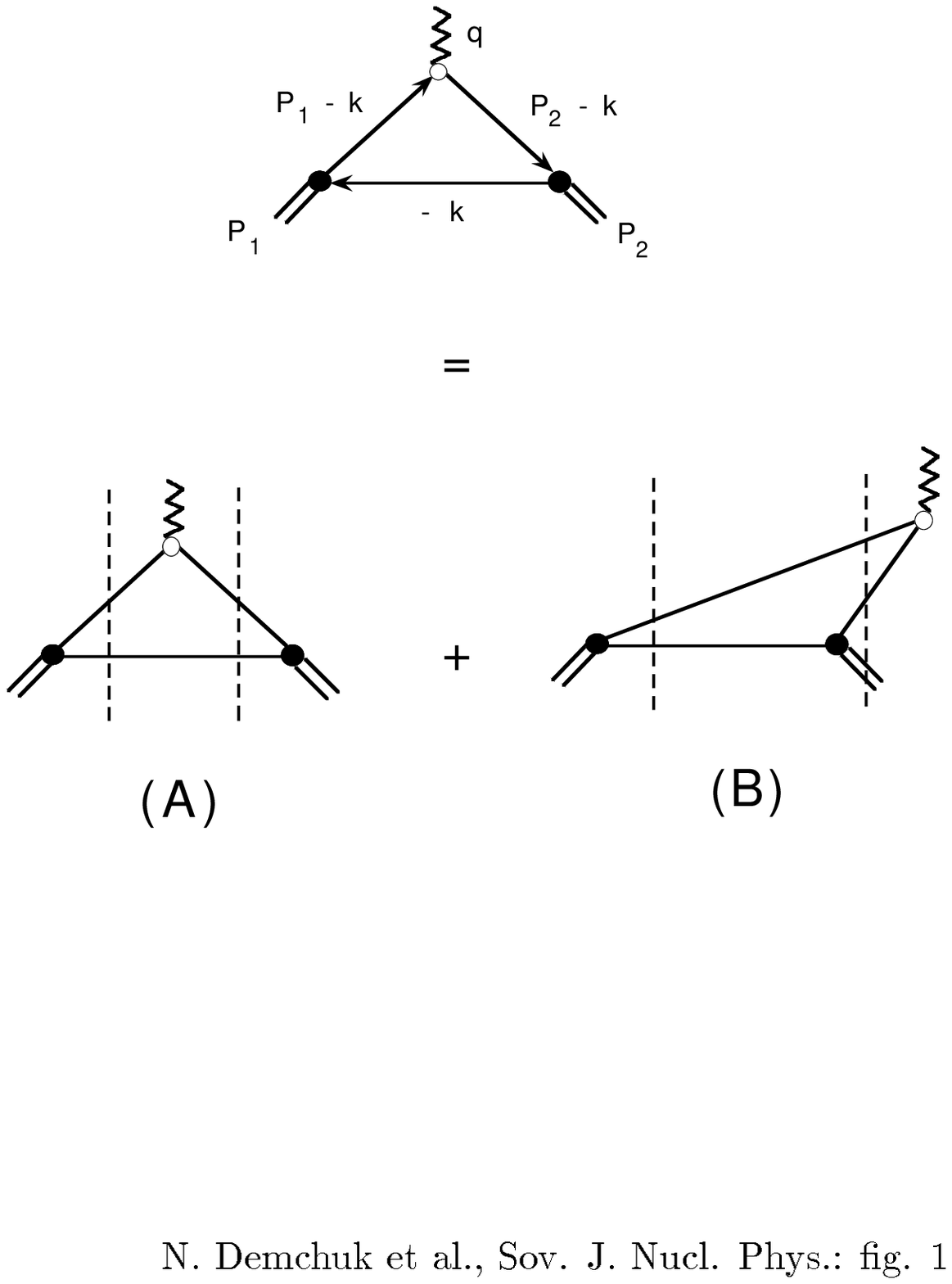}

\end{figure}

\newpage

\begin{figure}

\epsfig{file=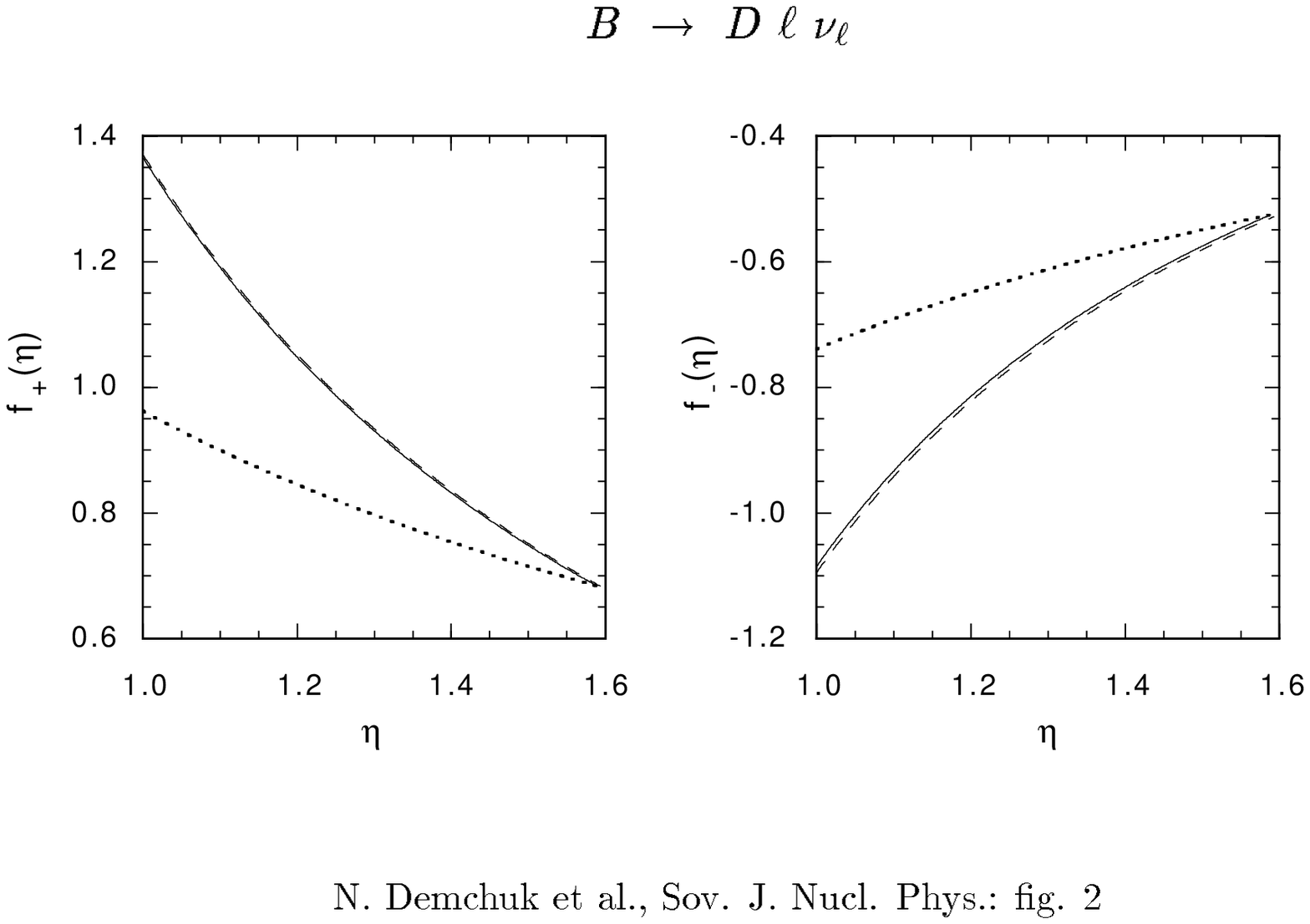}

\end{figure}

\newpage

\begin{figure}

\epsfig{file=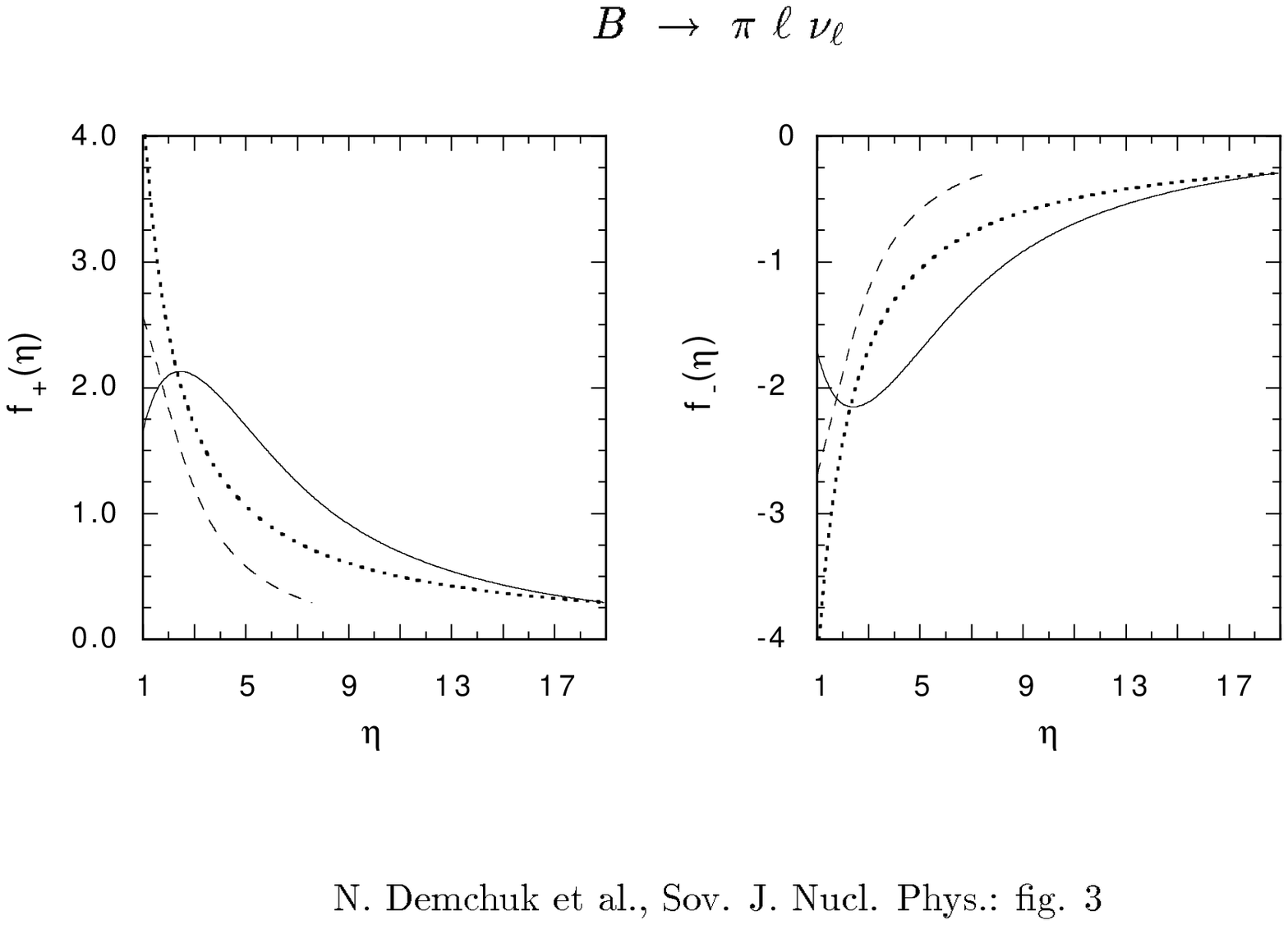}

\end{figure}

\newpage

\begin{figure}

\epsfig{file=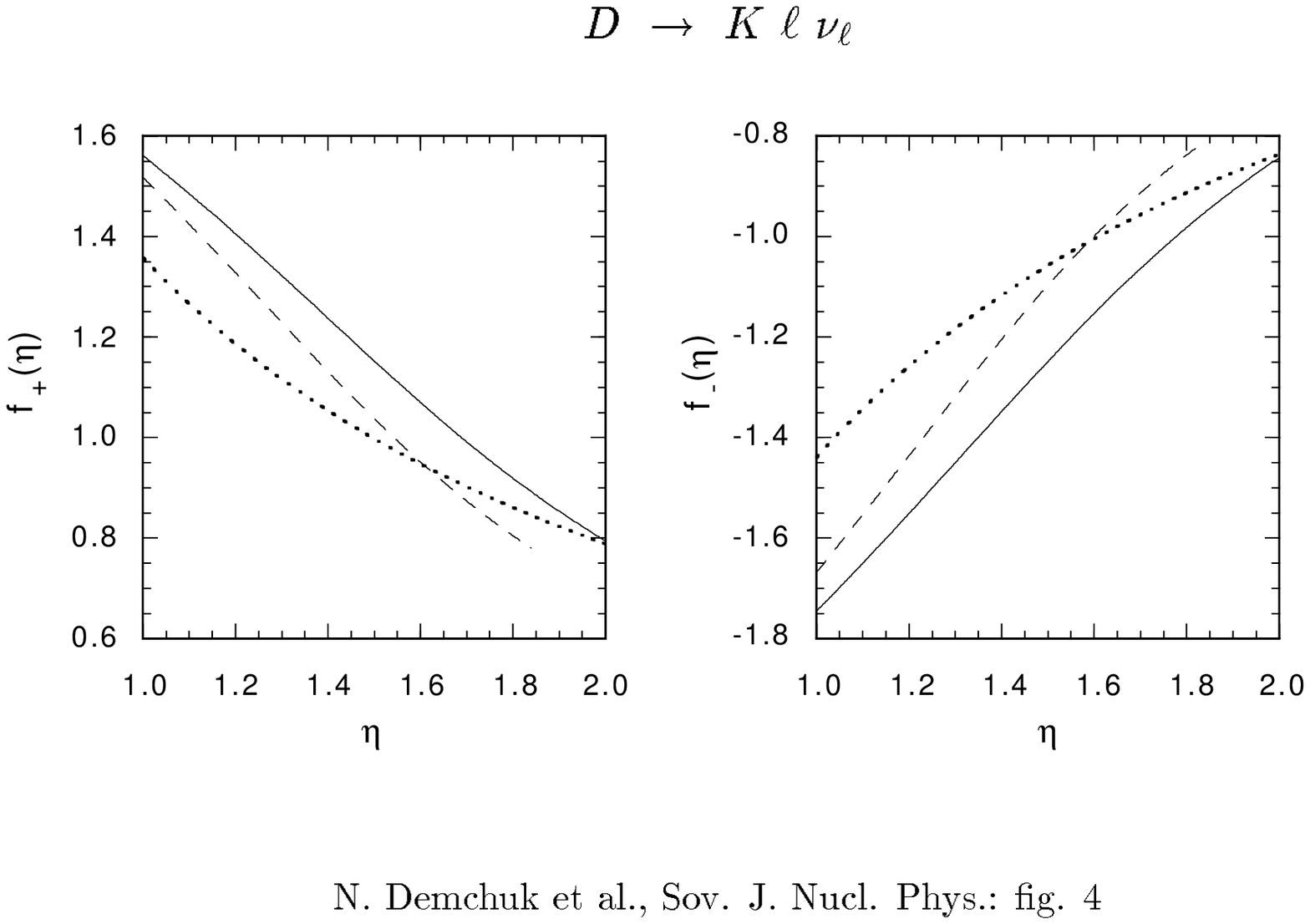}

\end{figure}

\newpage

\begin{figure}

\epsfig{file=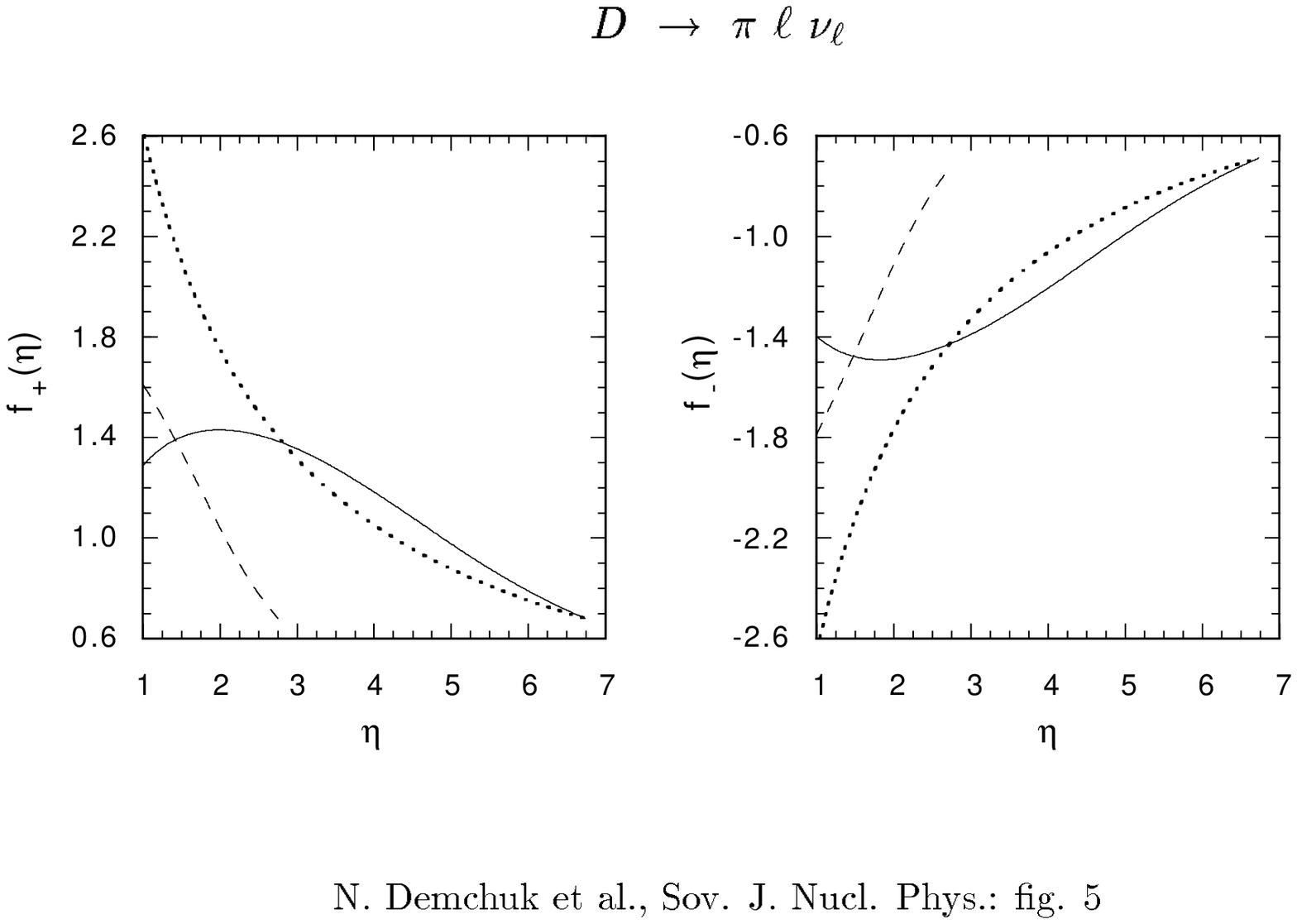}

\end{figure}

\newpage

\begin{figure}

\epsfig{file=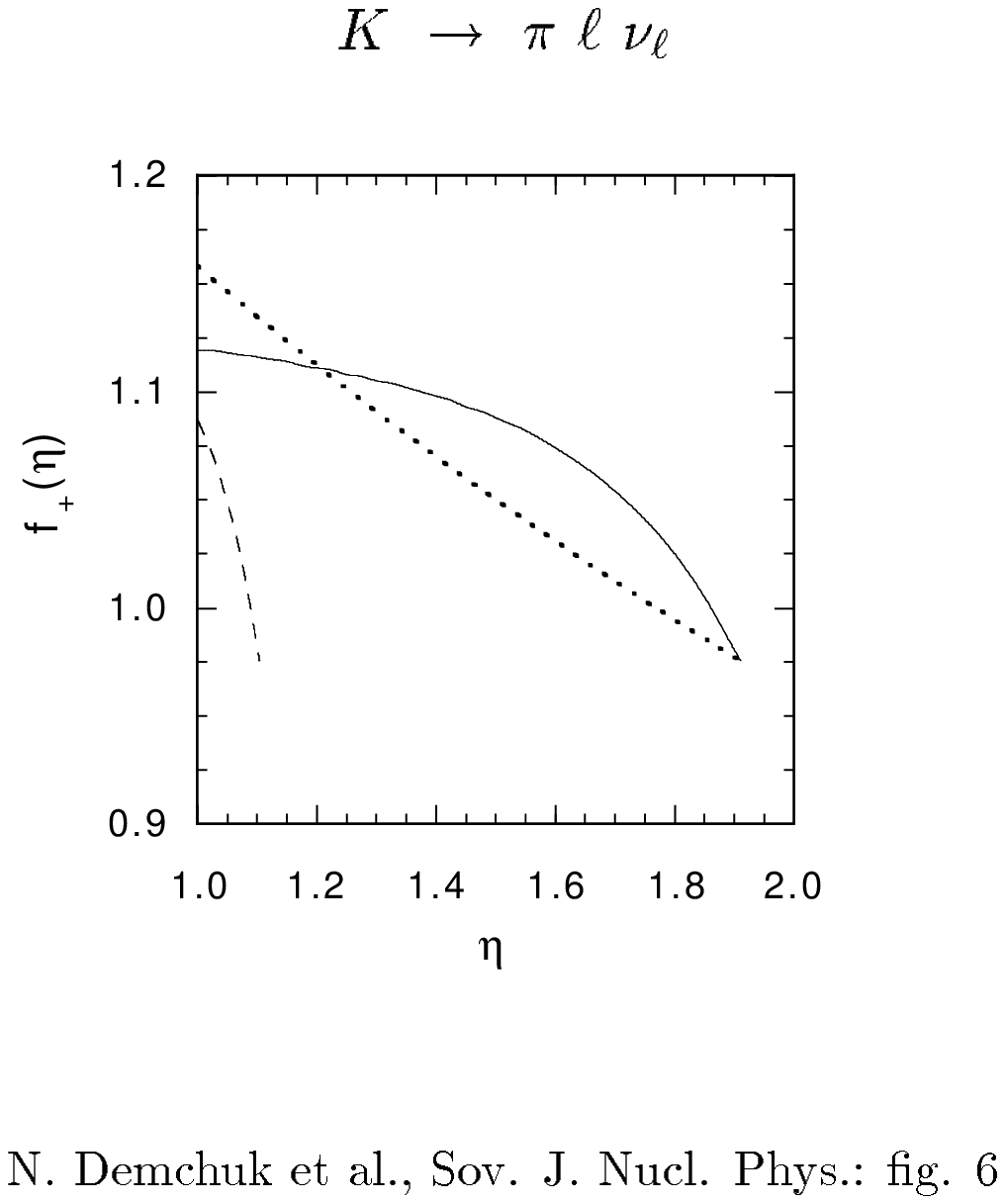}

\end{figure}

\end{document}